# EVALUATION OF TWO-LEVEL GLOBAL LOAD BALANCING FRAMEWORK IN CLOUD ENVIRONMENT

Po-Huei Liang[1] and Jiann-Min Yang [2]

Department of Management Information Systems, National Cheng Chi University, Taipei, Taiwan

## ABSTRACT

*With technological advancements and constant changes of Internet, cloud computing has been today's trend. With the lower cost and convenience of cloud computing services, users have increasingly put their Web resources and information in the cloud environment. The availability and reliability of the client systems will become increasingly important. Today cloud applications slightest interruption, the impact will be significant for users. It is an important issue that how to ensure reliability and stability of the cloud sites. Load balancing would be one good solution.*

*This paper presents a framework for global server load balancing of the Web sites in a cloud with two-level load balancing model. The proposed framework is intended for adapting an open-source load-balancing system and the framework allows the network service provider to deploy a load balancer in different data centers dynamically while the customers need more load balancers for increasing the availability.*

## KEYWORDS

*Network Load Balancing, Load Balancing as a Service, Cloud Computing, Virtualization.*

## 1. INTRODUCTION

Since 2009, Cloud Computing technology has become one of the hottest topics and different organizations have expressed a high degree of development of cloud computing attention and actively carried out relevant research and promotion.

Cloud computing is an Internet-based, according to the needs of the users, to provide sharing resources as computation computer, software information and other peripherals. And its architecture is based on the Internet that provides dynamic scalability for the consumer to use virtual resource-based information technology services.

As cloud computing provides dynamic scalability, flexibility, and virtualization resources, cloud computing applications are considered as an innovative optical department. What is the so-called cloud computing? According to the U.S. NIST (National Institute of Standards and Technology), the objective definition of cloud computing is a convenient way, according to the demand, to share networking, storage, applications, services, modules and other resources. Sharing of resources allows the providers can effectively release the extra resources to provide more effective services.

Load balancing is a way to spread requests out over multiple resources and it helps a network avoid annoying downtime and delivers optimal performance to users [6]. Network Load Balancing (NLB) can use a distributed algorithm to load balance network traffic across a number of hosts, helping to enhance the scalability as shown in figure 1. The global server load balancing (GSLB) can operate the Web site or another application server farm at multiple data centers and provide continuous availability by directing users to an alternative site when one site fails or the entire data center is down [8]. General server load balancing only limited to a single data center or near conduct, but GSLB can across different regions. Because the cost of hardware-based GSLB is usually expense, we would find out an economic solution in this paper.

With lower cost and more availability of cloud services, users have increasingly moved their applications to the cloud environment. Consequently, users can access their resources with the browsers of their thin devices, and cloud service providers do not need to buy numerous machines for the uncertain requirements of backup or expansion. Thus we would propose a software-based load balancing solution with cloud technology, this cost is much cheaper than hardware-based one.

The purpose of this paper is to present an open-source solution to build a two-level global load-balancing architecture that is scalable for Web clusters in cloud environments. The rest of this paper is organized as follows: Section II gives an overview of load balancing, the features and advantages of cloud computing are described in Section III. Section IV presents the proposed framework and lists the test results of the proposed framework. Finally, Section V discusses the conclusions and future research work.

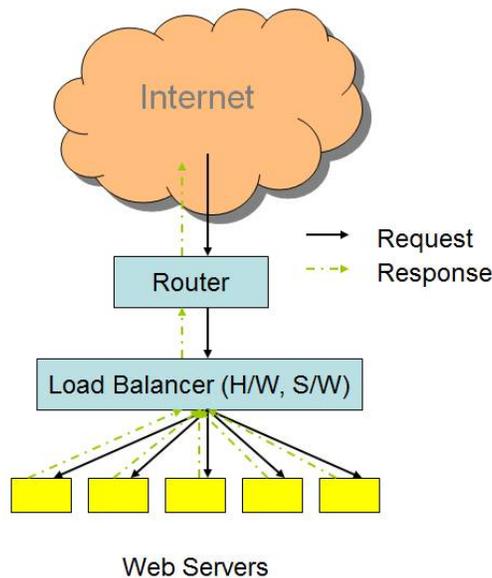

Figure 1. Workflow of network load balancing

## 2. OVERVIEW OF LOAD BALANCING

The most common problem faced by the Internet web provider is to resolve the congestion problem caused by the large amount of on-line users at the specific time interval. For a web administrator, the most and major improved problem is how to make web performance better quickly and provide better services for the users. There are many load balancing solutions which have been proposed and used to overcome these issues. Now there are many open source solutions, such as Linux Virtual Server (LVS) [22] that can be used to build a load balancing architecture. In order to solve various load problems, there are many known load balancing algorithms developed, which can be dived into two major categories [20, 41, 14, 30, 31] and dynamic [4, 9, 10, 32, 38]:

- Static load balancing policy: The scheduling of the static algorithm is carried out according to a predetermined approach. Static load balancing policies are generally based on the information about the average behavior of system; transfer decisions are independent of the actual current system state. Static load balancing schemes use a priori knowledge of the applications and statistical information about the system.

- Dynamic load balancing policy: It adapts its decision to the current state of the system. Therefore, the dynamic approach is more flexible to changes in system parameters. The choice of a load balancing algorithm is a difficult task. Thus many kinds of algorithms have been proposed and each of them varies based on specific application requirements.

Besides, there are many researches for the load balancing algorithm of some specific application [33, 34, 36, 39, 43]. Sometimes users would like to customize the configuration of their load balancers for optimizing services. On the other hand, the platform types of the load balancers can be divided into "hardware-based" and "software-based". The hardware-based solutions adapt the physical resources to build the load balancing environment, such as multilayer switches [11]. In this paper, we choose the "LVS" for the software-based load balancing system of proposed solution.

Table 1. Current status of LBaaS providers

| | Load balancing algorithms | Communication interface |
|---|---|---|
| **Amazon ELB** | • Round robin LB<br>• Sticky-session | HTTP, HTTPS, TCP, SSL, or Custom CLI |
| **Windows Azure** | • Performance LB<br>• Failover LB<br>• Round Robin LB | HTTP, HTTPS, TCP, UDP, SSL |
| **Rackspace** | • Random LB<br>• Round robin, weighted round robin weighted least connections<br>• Least connections, | HTTP, HTTPS, SSL, REST, TCP, UDP, LDAP, LDAPS, FTP, SFTP, POP3, SMTP |
| **HP CLB** | • Round robin<br>• Least connection | HTTP, HTTPS, TCP, REST, Python CLI, or HP Console UI |
| **GoGrid** | • Round Robin, weighted round robin<br>• Least connection (LC), weighted LC<br>• Source Address Hashing | HTTP, TCP, SSL |

Traditionally load balancers have been delivered to the client as a hardware product for balancing network traffic similar to typical computing hardware devices. The evolution of service oriented models such as Load Balancer as a Service (LBaaS) [28], has given the service vendors the opportunity to concentrate solely to their core capability. Thus instead of building data centers and infrastructure to deliver their product "load balancer", network service vendors started moving their deployment infrastructure to the cloud environment. In order to facilitate the load distribution among the rented server instances by a cloud tenant, the cloud provider has developed a business model where a load balancer an also be rented to the tenants similar to SaaS [28]. Currently load balancing as a service is provided by some popular cloud service providers shown in Table 1 [15-19, 21, 24, 26]. But LBaaS is still in its infancy and many problems still remain to be solved. The major needs and challenges of LBaaS are listed below [28]:

- Fault tolerance capability: Load balancer needs to have high fault tolerance capability.

- Elastic scalability: Elasticity is the most important feature for provided services through cloud computing; and the automatic deployment of resources is crucial and challenging.
- Network topology independence: The selected algorithm is highly dependent on the composition and the topology of the network of servers.

## 3. CLOUD COMPUTING

Cloud computing is the new evolution of on-demand information technology services and products [1]. Based on recent researches of cloud computing [2, 13, 23, 25, 37, 42], we find that there are several characteristics are the same and can be implemented as a service:

- On-demand self-service. Consumers can provision required computing capabilities automatically with service's providers.
- Ubiquitous network access. Customers can use device over the network to access the service.
- Location-independent resource pooling. Customers do not need to know the exact location of the services.
- Rapid elasticity. After the service purchased, it can be rapidly and elastically provisioned to quickly scale up, and rapidly released to quickly scale down.
- Pay per use. Fees are charged while using the demand service.

Based on cloud deployment approaches and community relations between clouds, there are several cloud models described in Table 2 [2, 23]. Beware of the type of deployment model instance in these cloud models would be internal or external.

Internet, hardware and systems software, operation management and other relative resources are included in the applied scopes of cloud services [12]. The resources of cloud computing will be provided in the virtualized type [1]. With the advantages of the resource virtualization, cloud service providers can dynamically "provision" on demand as a personalized resource collection to meet a specific service-level agreement [7, 29].

Table 2. Cloud deployment models

| Cloud model | Description |
|---|---|
| Private cloud | The cloud infrastructure is owned and operated solely for the specific organization. |
| Community cloud | The cloud infrastructure is shared by several organizations and supports a specific community with their agreed rules. |
| Public cloud | The cloud infrastructure is owned by an organization providing services to the general public customers. |
| Hybrid cloud | The cloud infrastructure is a composition of two or more clouds (private, community, or public). |

Cloud computing delivers resources as services, which in industry are respectively referred to as Infrastructure as a Service (IaaS), Platform as a Service (PaaS), and Software as a Service (SaaS) [2, 7, 23, 27, 29]. If the resources or the operation works of data centers are divided into more parts with single purpose, more innovative service categories can be generated, such as storage as a service, database as a service, testing as a service, etc [23].

## 4. TWO-LEVEL LOAD BALANCING FRAMEWORK

With the advantages of the lower cost and the dynamic scalability from cloud service, we implement the two-level global load balancing framework in the cloud environment. The main components of the proposed framework include the Load Balancer Selector (LBS) and the software-based load balancer. In order to implement a cloud solution, each component should be one or groups of virtual machines. To upgrade the high availability of the proposed framework, at least two of the LBS VMs include all load balancer IPs information and the LBS IP must be registered to the global DNS service provider. The load balancing system is to receive the http requests and then redirect them to the web system. It could be a single VM or a cluster for the high availability purpose in a cloud environment. If some VMs of two components need to be closed for the system maintenance purpose, the new alternative VMs can be deployed through requests. In the framework, users do not need prepare any hardware machines, network environments and the IT staffs.

In our proposed framework, we adapt the software-based load balancer to the Linux Virtual Server (LVS) [26]. The LVS can provide several techniques, such as network address translation, direct routing, and IP encapsulation, to distribute IP packets among nodes. Our proposed system chooses direct routing for sharing requests to different web servers in the cloud environment. The LVS is a virtual machine in the cloud and is supported only for one specific Web cluster in the same cloud service.

The two-level global load balancer proposed in this paper is an open-source load-balancing solution for Web clusters. This framework can be applied with other kinds of hypervisors, which can deploy new LVS VM instances, as shown in Figure 2. The LVS VM is responsible for balancing the loads from user requests. The LBS receives and arranges the HTTP requests from the users and transfers the links to the corresponding LVS addresses. Then, the corresponding LVS redirects the requests to the real Web server, which returns the HTTP results to the user. The purpose of this research was to develop an open-source solution that can rapidly be reused in the cloud environment, since the costs of these virtual load balancers are much less than those of the customary physical load balancers.

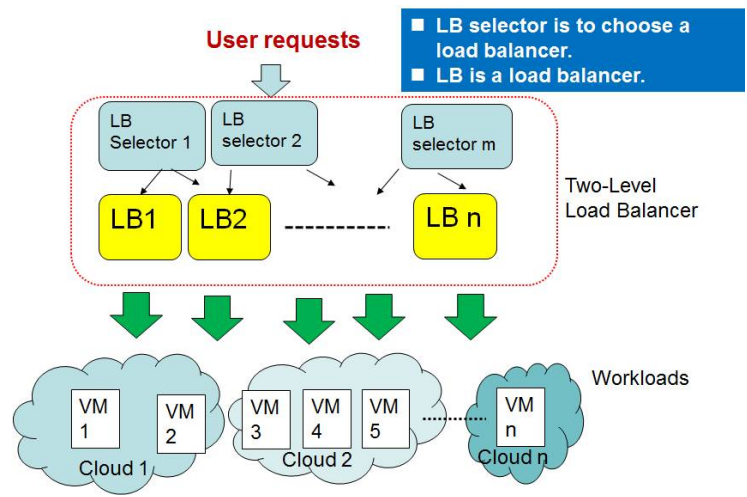

Figure 2. Architecture of two-level global load balancer

## 4.1. Proposed Architecture

Figure 2 shows the architecture of the proposed two-level global load balancer in the cloud environment. This framework includes the load balancer selector (LBS) and the load-balancing VMs in cloud environments. Each Web cluster runs with its own network load balancer (LVS), and the LVS systems are set up for high availability of a specific Web service. The LBS VM is responsible for sequentially arranging user requests to the corresponding load balancer VM IP address of the Web server cluster, and this function can spread the load of the first Web cluster. However, the second LBS VM becomes the slave of the LBS. In this architecture, the LBS is implemented by Round Robin Domain Name Server (RRNS) and the load balancer is adapted with Linux LVS.

In order to reduce the cost of GSLB and enhance the flexibility of load balancer, we proposed a kind of two-level global load balancing framework. Its information process can be described as the Figure 3. When the "ISP DNS" receives the AP1 request from the user1, the ISP DNS would check its DNS registry database and find the registered DNS and then transfers the http requests to the "LBS 1". The "LBS 1" would resolve the IP information and send to user1. Therefore the user1 would get the right IP to connect the load balancing system "LB VM 1", the "LB VM 1" would redirect the request to the "AP1 VM1". Finally, the user1 would get web data. In another case, there are two LBS VMs and two load balancer VMs for load balancing AP2 requests. When the "ISP DNS" receives the AP2 request from the user2, the ISP DNS would check its DNS registry database and find the registered DNS and then transfers the http requests to the "LBS 2". The "LBS 2" would resolve the IP information and send to user2. Therefore the user2 would get the right IP to connect the load balancing system "LB VM 2", the "LB VM 2" would redirect the request to the "AP2 VM1". Then the next AP2 request will redirect to "LB VM3" and user will get data from "AP2 VM4".

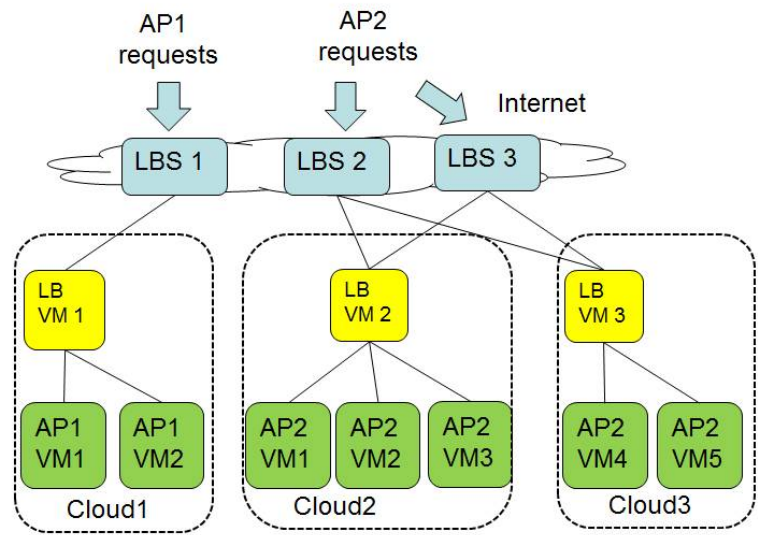

Figure 3. Workflow of two-level global load balancer

## 4.2. Implementation and Evaluation

The proposed framework was established and then evaluated. In the experimental environment, we adapted nine virtual machines created for the proposed framework. The first evaluation is to verify the high availability of the LBS. For the testing, we stop the major LBS VM "LBS VM01". We found that the http request can be resolved by the second LBS VM "LBS VM02" and return the web data correctly. The other evaluation is to validate the load balancing functionalities of the proposed framework. We make the different web contents on the five different web VMs, and then we use two isolated machines to connect the web system. The results of the two machines are different and this shows that our proposed framework is effective.

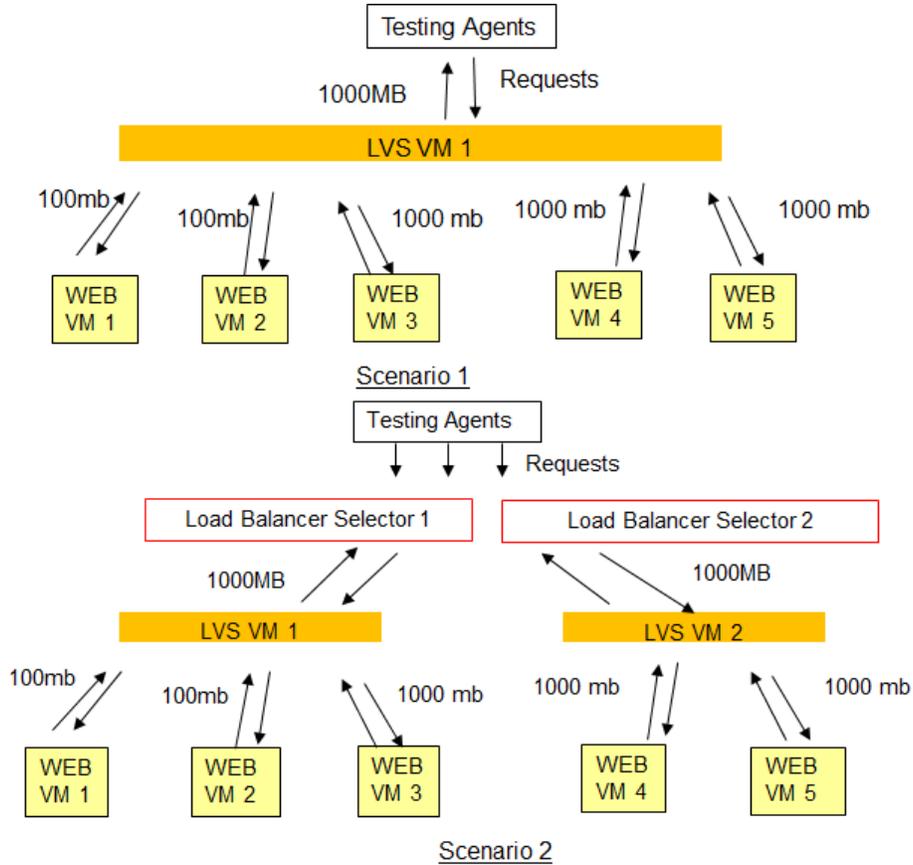

Figure 4. Testing environment of two experiment scenarios

The next demonstration explores the test results of the Web virtual machines in two different scenarios: 1) five Web VMs with one LVS load balancer, 2) five Web VMs with two load balancers (RRDNS and LVS). In our experiment, we use two HP ProLiant DL385 G6 servers, each containing one six-core 2.4-GHz AMD CPU with hyper-threading support and running the hypervisor of the Linux KVM [5]. Each machine is configured with 16 GB of memory and is connected with a single 1000 Mbit Internet bandwidth. The hardware configuration of all Web VMs is one vCPU and 1024 MB of RAM. The software-based load balancer is built with the Linux Virtual Server (LVS) on a CentOS Linux VM, and its configuration is one vCPU and 2048 MB of RAM. The RRDNS VM is established with one vCPU and 2048 MB of RAM. The environment of RRDNS is adopted with CentOS and BIND DNS.

Table 3. Testing results of evaluating response time

| Testing Scenarios | group 1 bandwidth of 3 web VMs (MB/s) | group 2 bandwidth of 2 web VMs (MB/s) | Load Balancing Algorithm | Testing results | | |
|---|---|---|---|---|---|---|
| | | | | Total Resp time (secs) | Avg Resp time (secs/req) | errors |
| (1) Five web VMs with one load balancer | 100/100/1000 | 1000/1000 | Round Robin | 90.889 | 0.454 | 0 |
| | 100/100/1000 | 1000/1000 | Weighted Least Connection | 94.116 | 0.471 | 0 |

| (2) Five web VMs with two-level global load balancer (two load balancers) | 100/100/1000 | 1000/1000 | Weighted Least Connection + Weighted Least Connection | 47.923 | 0.239 | 0 |
| | 100/100/1000 | 1000/1000 | Weighted Least Connection + Round Robin | 54.614 | 0.273 | 0 |
| | 100/100/1000 | 1000/1000 | Round Robin + Round Robin | 63.259 | 0.316 | 0 |

The load-generating client adopted is a Personal Computer with the Linux CentOS operating system installed. The hardware configuration of the client is one four-core CPU and 8192 MB of RAM. The Web stress tools used on the client are ApacheBench (ab) [3] and Web Bench [40], those tools are open-source Web performance testing tool. First experiment tested by ApacheBench is to evaluate the response time with different load balancing algorithms in two experiment scenarios. Each testing case executes 200 requests with 100 concurrent agents. In this study, every testing case were tested 30 times and the testing results were gathered and averaged in Table 3. Another experiment tested by Web Bench is to evaluate the finished requests with different load balancing algorithms in two experiment scenarios. All 100 agents run with the same configuration, which includes the duration of 300 s, the ramp-up time of 0 s, and the interval time of 0 ms. The stress testing is conducted with various agent parameters and the test results are gathered in Table 4. The results show that the requests generated by the proposed framework were much more the results of one load balancer.

Table 4. Testing results of evaluating finished requests

| Testing Scenarios | group 1 bandwidth of 3 web VMs | group 2 bandwidth of 2 web VMs | Load Balancing Algorithm | Results | |
| --- | --- | --- | --- | --- | --- |
| | | | | Total Requests | Throughput |
| (1) Five web VMs with one load balancer | 100/100/1000 | 1000/1000 | Round Robin | 655 | 131pages/min |
| | | | Weighted Least Connection | 608 | 121 pages/min |
| (2) Five web VMs with two-level global load balancer (two load balancers) | | | Weighted Least Connection + Weighted Least Connection | 721 | 144 pages/min |
| | | | Weighted Least Connection + Round Robin | 673 | 137pages/min |
| | | | Round Robin + Round Robin | 707 | 141 pages/min |

## 5. CONCLUSIONS

This study is focused on the study is to verify and improve the framework of virtualized network load balancers. The proposed framework is expected to be replaced with the physical network load balancers, which are expensive. Based on the benefits of Cloud Computing and Virtualization Technology, we believe that the proposed framework in this study could save the IT cost and be deployed rapidly. At last, we proposed an executable two-level load balancing model with two different algorithms and implemented some experiment to validate its function. Strictly speaking, the expected contribution of this study are as follows:

The general solution for solving the high availability of the web applications is to connect with the network load balancer. Although the performance of the hardware-type network load balancer is higher than the one of the software-type network load balancer. But its cost is many times higher than the cost of the software-type. It makes that enterprise users would not buy or expand new machines easily.

In practice, we constructed an architecture of two-level global load balancer in this study and the architecture is open source and low-cost. We further integrated with the features of cloud computing and established the virtualized network load balancers in different cloud environment. This service would be used to solve the web connection limitation of the network load balancer in a single network. It could be realized in the enterprise environment and the Internet Data Center, and the IT staffs could implement the service easily. In addition, it allows users themselves to update the load balancing algorithms of load balancers dynamically in different cloud environments. Further, update the load balancer selector while adding or decrease the amount of the load balancers.

Although the proposed concept has been proved, many aspects need to be improved and compared with those of other load-balancing algorithms. Our future work will focus on how to improve the performance for large users over the Internet and make comparisons with other kinds of load-balancing algorithms. Besides, we would conduct more comprehensive evaluations on our system in hybrid cloud environment.

## REFERENCES

[1]  A Vouk, M. (2008). Cloud computing–issues, research and implementations. CIT. Journal of Computing and Information Technology, 16(4), 235-246.
[2]  Mell, P., & Grance, T. (2009). The NIST definition of cloud computing. National Institute of Standards and Technology, 53(6), 50.
[3]  ApacheBench (ab), http://en.wikipedia.org/wiki/ApacheBench
[4]  Arora, M., Das, S. K., & Biswas, R. (2002). A de-centralized scheduling and load balancing algorithm for heterogeneous grid environments. In Parallel Processing Workshops, 2002. Proceedings. International Conference on (pp. 499-505). IEEE.
[5]  Kernel-based Virtual Machine (KVM), http://www.linux-kvm.org/page/Main_Page.
[6]  Mishra, M. A. Network Load Balancing and Its Performance Measures.
[7]  Buyya, R., Ranjan, R., & Calheiros, R. N. (2009, June). Modeling and simulation of scalable Cloud computing environments and the CloudSim toolkit: Challenges and opportunities. In High Performance Computing & Simulation, 2009. HPCS'09. International Conference on (pp. 1-11). IEEE.
[8]  Kopparapu, C. (2002). Load balancing servers, firewalls, and caches. John Wiley & Sons.
[9]  Dhakal, S., Hayat, M. M., Pezoa, J. E., Yang, C., & Bader, D. A. (2007). Dynamic load balancing in distributed systems in the presence of delays: A regeneration-theory approach. Parallel and Distributed Systems, IEEE Transactions on, 18(4), 485-497.
[10] Dobber, M., Koole, G., & van der Mei, R. (2005, May). Dynamic load balancing experiments in a grid. In Cluster Computing and the Grid, 2005. CCGrid 2005. IEEE International Symposium on (Vol. 2, pp. 1063-1070). IEEE.

[11] Load Balancing, http://en.wikipedia.org/wiki/Load_balancing_%28computing%29.

[12] Fox, A., Griffith, R., Joseph, A., Katz, R., Konwinski, A., Lee, G. & Stoica, I. (2009). Above the clouds: A Berkeley view of cloud computing. Dept. Electrical Eng. and Comput. Sciences, University of California, Berkeley, Rep. UCB/EECS, 28, 13.

[13] Greer, M. B. (2009). Software as a service inflection point: Using cloud computing to achieve business agility. iUniverse.

[14] Grosu, D., & Chronopoulos, A. T. (2005). Noncooperative load balancing in distributed systems. Journal of Parallel and Distributed Computing, 65(9), 1022-1034.

[15] Amazon White Paper, Elastic Load Balancing Developer Guide [Online]. Available: http://awsdocs.s3.amazonaws.com/ElasticLoadBalancing/latest/elb-dg.pdf.

[16] Elastic Load Balancing. Available [Online]: https://aws.amazon.com/elasticloadbalancing/.

[17] About Load Balancing Methods [Online]. Available: http://msdn.microsoft.com/en-us/library/windowsazure/dn339010.aspx.

[18] Windows Azure Traffic Manager [Online], Available: http://msdn.microsoft.com/en-us/library/windowsazure/hh744833.aspx.

[19] Rackspace Cloud Load Balancers. Available [Online]:http://www.rackspace.com/cloud/load-balancing/.

[20] Kameda, H., Li, J., Kim, C., & Zhang, Y. (2011). Optimal load balancing in distributed computer systems. Springer Publishing Company, Incorporated.

[21] The Technologies Behind Cloud Load Balancers. Available [Online]: http://www.rackspace.com/cloud/load-balancing/technology/.

[22] Linux Virtual Server (LVS), http://www.linuxvirtualserver.org/.

[23] Linthicum, D. S. (2009). Cloud computing and SOA convergence in your enterprise: a step-by-step guide. Pearson Education.

[24] HP Cloud Load Balancer. Available [Online]:http://www.hpcloud.com/products-services/load-balancer?t=features.

[25] Menken, I., & Blokdijk, G. (2010). Cloud Computing Virtualization Specialist Complete Certification Kit-Study Guide Book and Online Course. Emereo Pty Ltd.

[26] GoGrid Load Balancer. Available [Online]: http://www.gogrid.com/products/load-balancers.

[27] Motahari-Nezhad, H. R., Stephenson, B., & Singhal, S. (2009). Outsourcing business to cloud computing services: Opportunities and challenges. IEEE Internet Computing, 10.

[28] Rahman, M., Iqbal, S., & Gao, J. (2014, April). Load Balancer as a Service in Cloud Computing. In Service Oriented System Engineering (SOSE), 2014 IEEE 8th International Symposium on (pp. 204-211). IEEE.

[29] Youseff, L., Butrico, M., & Da Silva, D. (2008, November). Toward a unified ontology of cloud computing. In Grid Computing Environments Workshop, 2008. GCE'08 (pp. 1-10). IEEE.

[30] Penmatsa, S., & Chronopoulos, A. T. (2005, April). Job allocation schemes in computational Grids based on cost optimization. In Parallel and Distributed Processing Symposium, 2005. Proceedings. 19th IEEE International (pp. 180a-180a). IEEE.

[31] Penmatsa, S., & Chronopoulos, A. T. (2006, April). Price-based user-optimal job allocation scheme for grid systems. In Parallel and Distributed Processing Symposium, 2006. IPDPS 2006. 20th International (pp. 8-pp). IEEE.

[32] Penmatsa, S., & Chronopoulos, A. T. (2007, March). Dynamic multi-user load balancing in distributed systems. In Parallel and Distributed Processing Symposium, 2007. IPDPS 2007. IEEE International (pp. 1-10). IEEE.

[33] Liu, Y., Wang, L., & Li, S. (2008, November). Research on self-adaptive load balancing in EJB clustering system. In Intelligent System and Knowledge Engineering, 2008. ISKE 2008. 3rd International Conference on (Vol. 1, pp. 1388-1392). IEEE.

[34] Lee, W., Lee, H. W., & Choi, M. (2013, October). Load balancing system for IPTV web application virtualization. In ICT Convergence (ICTC), 2013 International Conference on (pp. 602-603). IEEE.

[35] Rimal, B. P., Choi, E., & Lumb, I. (2009, August). A taxonomy and survey of cloud computing systems. In INC, IMS and IDC, 2009. NCM'09. Fifth International Joint Conference on (pp. 44-51). IEEE.

[36] Guo, J., & Bhuyan, L. N. (2006). Load balancing in a cluster-based web server for multimedia applications. Parallel and Distributed Systems, IEEE Transactions on, 17(11), 1321-1334.

[37] Rittinghouse, J. W., & Ransome, J. F. (2009). Cloud computing: implementation, management, and security. CRC press.

[38] Shah, R., Veeravalli, B., & Misra, M. (2007). On the design of adaptive and decentralized load

balancing algorithms with load estimation for computational grid environments. Parallel and Distributed Systems, IEEE Transactions on, 18(12), 1675-1686.

[39] Niyato, D., & Srinilta, C. (2001, October). Load balancing algorithms for internet video and audio server. In Networks, 2001. Proceedings. Ninth IEEE International Conference on (pp. 76-80). IEEE.

[40] Web Bench, https://xuri.me/2013/10/27/install-webbench.html.

[41] Tang, X., & Chanson, S. T. (2000). Optimizing static job scheduling in a network of heterogeneous computers. In Parallel Processing, 2000. Proceedings. 2000 International Conference on (pp. 373-382). IEEE.

[42] Velte, T., Velte, A., & Elsenpeter, R. (2009). Cloud computing, a practical approach. McGraw-Hill, Inc.

[43] Menon, H., & Kalé, L. (2013, November). A distributed dynamic load balancer for iterative applications. In Proceedings of SC13: International Conference for High Performance Computing, Networking, Storage and Analysis (p. 15). ACM.

.